# Photometric monitoring of the ROSAT selected weak-line T Tauri stars in the Taurus-Auriga field

Li-Feng Xing[1,2] ⋆, Xiao-Bin Zhang[1] and Jian-Yan Wei[1]

[1] National Astronomical Observatories, Chinese Academy of Sciences, Beijing 100012, China

[2] Graduate University of the Chinese Academy of Sciences, Beijing 100049, China

**Abstract** We monitored the light variations of 22 weak-line T Tauri stars (WTTS) discovered among the X-ray sources in the field of the Taurus-Auriga cloud. The photometric variability of 12 out of these WTTS samples is confirmed. They are all proved to be in periodic variations. By using the methods of the Phase Dispersion Minimization (PDM) and Fourier analysis, the rotational periods of these stars were determined based on this observation. Most of them are found to be shorter than one day. This gives further evidences for the spin up of solar-type stars as predicted by models of angular momentum evolution of pre-main sequence stars.

**Key words:** Stars: late-type– Stars: pre-main sequence– Stars: rotation

## 1 INTRODUCTION

T Tauri stars (TTSs) are young ($\leq 10^8$ yr), low-mass ($M \leq 2 M_\odot$), late spectral type (typically G0 or later) pre-main sequence (PMS) stars. It is well-known that TTSs are photometric variables (Joy 1945). These stars were originally classified as irregular variables (Herbig 1962). Their light variation periods span from minutes to decades and amplitudes from a few magnitudes down to a few hundredths (Herbst 1994). Irregular variations of up to several magnitudes are interpreted as a result of accretion and, perhaps, occultation events within the dusty and gaseous disks surrounding classical T Tauri stars (CTTSs). One also sees periodic variations of typically a few tenths of a magnitude or less in weak-line T Tauri stars (WTTSs) that may be largely or entirely attributed to cool (magnetic) spots on the stellar surface (Herbst 1994). Photosphere's spots modulate the light curves at the rotational period of the star and thus allows the rotational period to be obtained.

⋆ E-mail: `lfxing@bao.ac.cn`

Up to date, photometric surveys in star formation regions and young open clusters have resulted in several hundreds of periodic variation WTTSs and youngest dwarfs(Bouvier et al. 1993, 1995; Grankin et al. 1995; Prosser et al. 1995; Shevchenko et al. 1998; Herbst et al. 2000; Lawson et al. 2001). But the amount of short-period (<1d) samples is still very small. It is hard to understand that there seems to be a gap in periods between one day and a few hours, or to say, a blank between the oldest TTSs and the youngest variable dwarfs. According to the theory of stellar structure evolution, when a low-mass star evolves from PMS to normal main-sequence, its interior will change rapidly from completely convective to mostly radiative, and this usually causes the most rapid and drastic surface variations such as spot activity and chromospheric emissions. Therefore, searching for short-period WTTSs could be a subject of special interest. To do that, the X-ray sources in star formation regions must be the most suitable objects, since the targets as optical counterparts of X-ray sources that might have selected stars with very high coronal emission, typical of faster rotators.

The Taurus-Auriga cloud is a famous star formation region. In this field, the ROSAT all-sky survey has identified a large number of X-ray sources. Photometric observations on these X-ray sources revealed a total number of 22 WTTSs or WTTS candidates(O'Neal et al. 1990, Gregorio-Hetem & Hetem 1992, Wichmann et al. 1996, Li & Hu 1998). Rotational periods of 3 definite WTTSs were determined by Zakinov et al. (1993), Grankin (1994) and Bouvier et al. (1997), respectively. One of them was found to be a short-period variable. For the other 19 candidates, their light variation nature and periods are still uncertain. Among them, we predict that there could be some more short-period WTTS samples (most of the stars are found located outside of the dense dark cloud in which optical-selected classical T Tauri stars usually appear in clusters).

In the observation seasons of 2004 and 2005, we performed a long-term CCD photometric monitoring on all these 22 stars. Our main goal is to derive the rotational periods of these WTTS, and hope to search for more WTTS samples with period cover the gap as mentioned above. In this paper, we present the results of the observations. Based on which, a brief discussion about the angular momentum evolution of PMS stars will given.

## 2 OBSERVATIONS AND DATA REDUCTION

Table 1 lists the basic data of the program stars. In which, 13 samples are collected from Li & Hu (1998), 7 are adopted from Wichmann (1996), remainder both one is collected from O'Neal et al.(1990) and another from Gregorio-Hetem & Hetem (1992).

The observations were performed from October 2004 to February 2005 on the 60-cm reflector telescope at Xinglong Observatory of the National Astronomical Observatories, Chinese Academy of Science (NAOC) equipped with a PI $1300 \times 1340$ photometric CCD. The plate-scale of the camera is 0".46/pixel, providing a total field of view of $9'.93 \times 10'.24$. The standard Johnson B, V and R filters were used. In total, we obtained more than 200 CCD frames in each band for each star.

Data reduction is carried out with the Image Reductions and Analysis Facility (IRAF) software. All images are corrected for electronic bias and pixel-to-pixel gain variations (with sky flats). The routines of the DAOPHOT package are used to obtain relative magnitudes for the objects selected in each crowded field and the APPHOT package are used in each uncrowded field. We adopt the standard method to perform the differential photometry. In each field, we analyzed light-curves of several stars and selected those did not vary over the ten nights as our reference star. for these stars we have rejected those displaying instrumental magnitudes and colors very different from that of the targets. The remaining reference stars were examined for stability and we computed their averaged instrumental magnitudes as the magnitude of "artificial comparison". The transformation of the instrumental magnitudes to a standard system was not done, since our main interest is to study the light variations of star with time. The precision of differential photometry for all the stars is better than 0.023 mag.

## 3 RESULTS AND DISCUSSIONS

Our photometric observations confirm the variability of 12 out of the 22 program samples. The time-serious BVR light curves of these definite WTTSs are displayed in Fig.1. For the other 10 stars, we failed to detect any light variations under the present photometric accuracy. It is suggested that they might be WTTSs with low-amplitudes, long periods variable though we could not exclude that some of them might be long-period, low-amplitude variables.

With the newly derived photometric measurements, the rotational periods of the 12 variables were determined. The periods were computed by using the Phase Dispersion Minimization (PDM) method (Lafler & Kinman 1965, Stellingwerf 1978), and checked applying the Fourier analysis with the code PERIOD04 (Lenz & Breger 2004). In Table 2, we present the main results for the 12 WTTSs. The phased V-band light curves formed with the derived periods are shown in Fig. 2.

For the stars, HD 285840 and TAP 57NW, their periods determined by us are in good agreement with that given by Bouvier et al.( 1997, period of HD 285840 is 1.55 days) and Grankin ( 1994,period of TAP 57NW is 9.34 days), respectively. For HD 285372, its period was refined as 0.287 days, which is just a half of the value of 0.573 days previously reported by Bouvier et al. ( 1997).

Among the 12 definite variables, we find that most (seven) of them are short-period (with P<1 days, very near the periods of stars in Pleiades cluster) WTTSs. This confirms our prediction discussed above. Our result improves largely the WTTSs samples and provides important evidences in study of the angular momentum evolution of PMS stars.

The amplitude of the photometric variability might depend on the periods. In order to study this we have compiled from the literature data on Post-TTSs (Bouvier et al. 1997) and Zero-Age Main Sequence (ZAMS) clusters (Messina 2001). In Fig.3 we plotted the photometric amplitude ($\Delta V$) vs. rotational period for Post-T Tauri stars and ZAMS stars. There is indeed an obvious indication for the stars with cool spots that no dependency of amplitudes on periods is discernible within this range of rotational periods.

The figure of the rotational periods versus (B-V) color indices (Fig.4) does not show any particular trend in the spectral range from G0 to K7. This behavior confirms what already found for young open clusters(e.g. Marilli et al. 1997).

The rotational periods is plotted versus the $H_\alpha$ equivalent for the G0 to K7 sample in Fig.5. The figure shows that there is no correlation to be found between EW($H_\alpha$) and $P_{rot}$ within WTTSs and Post- T Tauri stars.

In order to investigate the possible effect of rotation upon Lithium depletion, we plotted in Fig.6 the Li I $\lambda$ 6707 Å equivalent versus the rotational periods for stars of our sample and the WTTSs of Bouvier et al. ( 1993) sample. Only stars with a spectral type between G0...K7 have been consider. There is a hint of a correlation between EW(Li) and $P_{rot}$ in the sense that, on average, slow rotors have higher lithium equivalent width than fast rotation WTTSs (Bouvier et al. 1993). Furthermore, when the rotational periods down to about three days or less the Li depletion staring hasten. Such a correlation may indicate that rapid rotation lead to a faster lithium depletion in WTTS phase. It needs more observations for obtained the rotational periods and equivalent widths of WTTSs and further confirm.

## 4 CONCLUSIONS

In summary, we have present the photometric monitoring of 22 X-ray emission WTTSs in the Taurus-Auriga SFRs. The results confirms the variability of 12 out of these WTTSs. They are all in periodic variations. Based on the new measurements, the rotational periods of these 12 variables are determined. They range from 0.287 to 9.12 days. Several models have been proposed in the last years to account for the surface rotation rates evolution for solar-type stars from their youngest T Tauri stage up to the age of ZAMS (e.g. Bouvier 1994; Keppens et al. 1995; Bouvier 1997 and Collier Cameron et al. 1995). The results report here provide further evidences for enhanced angular velocity in PMS stars approaching the ZAMS on their radiative tracks as predicted by models of angular momentum evolution of pre-main sequence stars.

**Acknowledgements** The authors would like to thank referee for valuable suggestions. We also would like to thank Prof. Jing-Yao Hu for the valuable discussion. The authors would also like to acknowledge Dr. Cai-Na Hao and Dr. Feng-Shan Liu for help of doing Monte-Carlo simulations, language and the valuable discussion. We thank Dr. Jing Wang for discussion on the data reduction and help of language.

---

This manuscript was prepared with the ChJAA LaTeX macro v1.0.

**Table .1** The sample stars of our observations. References to table 1: L: Li & Hu ( 1998), W: Wichmann ( 1996), O: O'Neal et al. ( 1990), G: Gregorio-Heterm et al. ( 1992)

| Objects | RA($\alpha$2000) | DEC($\delta$2000) | SpT | B(mag) | V(mag) | EW(Li)(mÅ) | EW(H$\alpha$)(Å) | Class. | ref. |
|---|---|---|---|---|---|---|---|---|---|
| 1 | 2 | 3 | 4 | 5 | 6 | 7 | 8 | 9 | 10 |
| [LH98 ]37 | 03:03:57 | 37:39:05 | K0IV | | 11.7 | -240 | 1.01 | WTTS | L |
| [LH98 ]53 | 03:16:43 | 19:23:04 | G0 | 11.33 | 11.07 | -230 | 0.76 | WTTS | L |
| [LH98 ]56 | 03:19:07 | 39:34:10 | K0V | | 11.6 | -300 | 0.68 | WTTS | L |
| [FS2003 ]0127 | 03:25:48 | 36:51:47 | K0IV | | 13.1 | -270 | 0.94 | WTTS | L |
| [LH98 ]87 | 03:44:12 | 24:01:54 | K0V | 12.26 | 10.86 | -210 | 1.44 | WTTS | L |
| [LH98 ]98 | 03:46:29 | 24:26:05 | K0V | | 11.4 | -190 | 0.71 | WTTS | L |
| HD 285281 | 04:00:31 | 19:35:20 | K0 | 11.17 | 10.4 | -258 | | WTTS | W |
| HD 285372 | 04:03:24 | 17:24:26 | K3 | 12.77 | 11.73 | -487 | | PTTS | W |
| HD 283323 | 04:05:12 | 26:32:44 | K2 | 12.29 | 11.47 | -210 | | WTTS | W |
| HD 26182 | 04:10:04 | 36:39:12 | G0 | 9.99 | 9.47 | -140 | 1.00 | WTTS | L |
| HD 284503 | 04:30:49 | 21:14:10 | G8 | 11.20 | 10.30 | -141 | | WTTS | W |
| HD 285840 | 04:32:42 | 18:55:09 | K1 | 10.78 | 10.8v | -253 | | WTTS | W |
| HD 283716 | 04:34:39 | 25:01:01 | K0IV | 11.10 | 10.33 | -50 | | WTTS | O |
| HD 282346 | 04:39:31 | 34:07:46 | K2 | 10.42 | 9.65 | -240 | 0.96 | WTTS | L |
| GSC 01292-00639 | 04:50:00 | 22:29:57 | K1 | | 11.08 | -275 | | WTTS | W |
| TAP 57NW | 04:56:02 | 30:21:03 | K5 | 12.88 | 11.60 | -580 | | WTTS | W |
| [LH98 ]179 | 05:15:49 | 18:44:20 | G7IV | 11.4 | 10.7 | | | WTTS | L |
| HD 287927 | 05:30:48 | 02:59:34 | G5 | 11.30 | 10.6 | -200 | 0.5 | WTTS | G |
| HD 244354 | 05:31:04 | 23:12:34 | G0 | 9.78 | 9.18 | -350 | 1.78 | WTTS | L |
| [LH98 ]212 | 05:36:50 | 13:37:56 | K0V | 11.4 | 10.7 | -360 | 1.63 | WTTS | L |
| HD 245358 | 05:36:51 | 23:26:15 | G0 | 9.48 | 8.8 | -300 | 1.24 | WTTS | L |
| HD 245567 | 05:37:18 | 13:34:52 | G5 | 10.26 | 9.53 | -250 | 0.80 | WTTS | L |

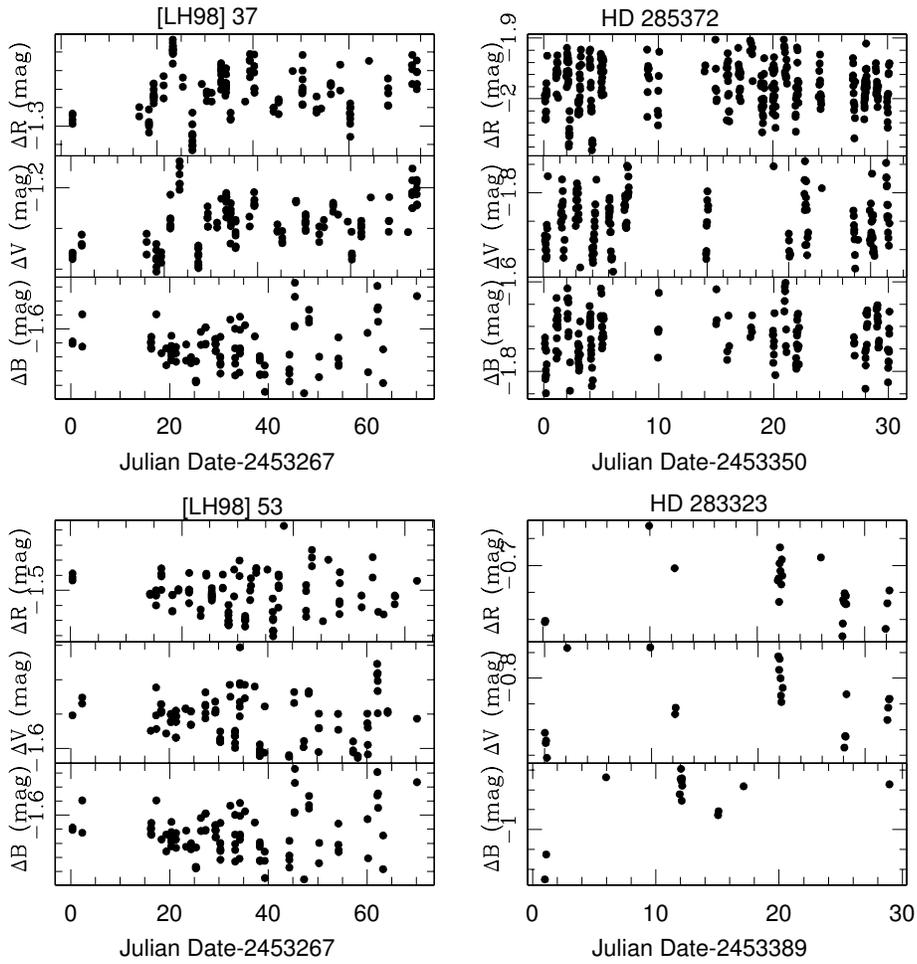

**Fig. .1** BVR light curves of the observed WTTSs

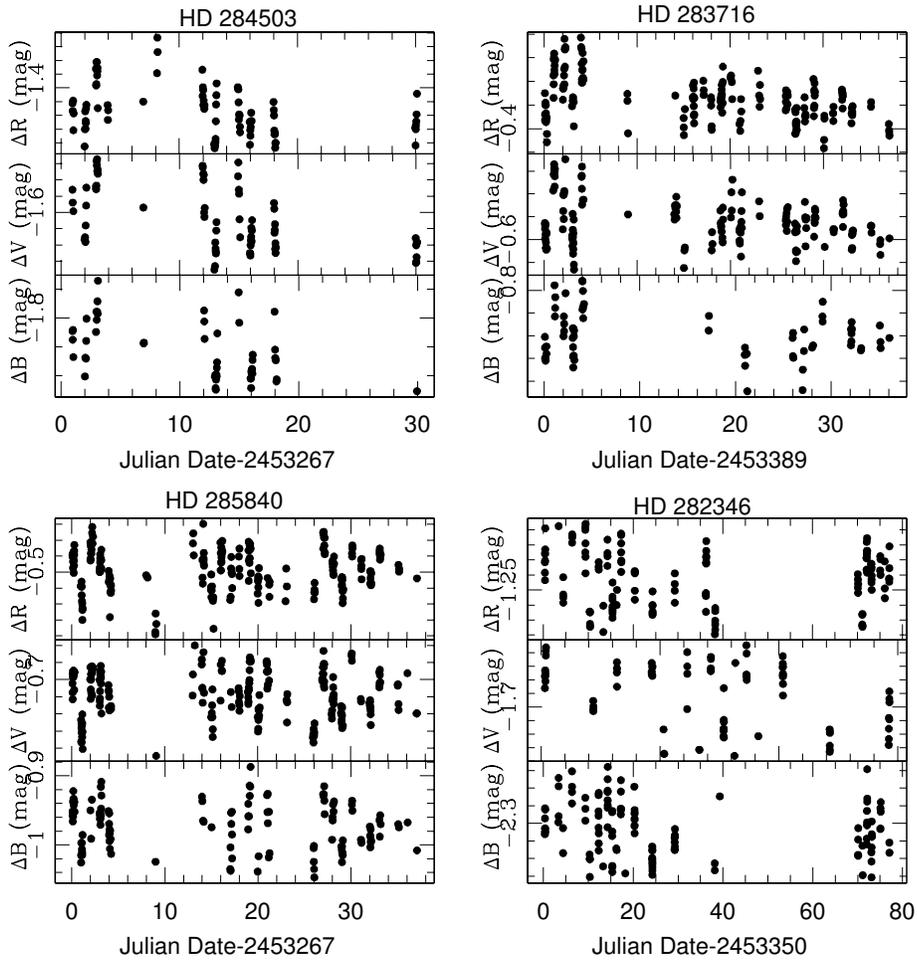

**Fig. .1** continued

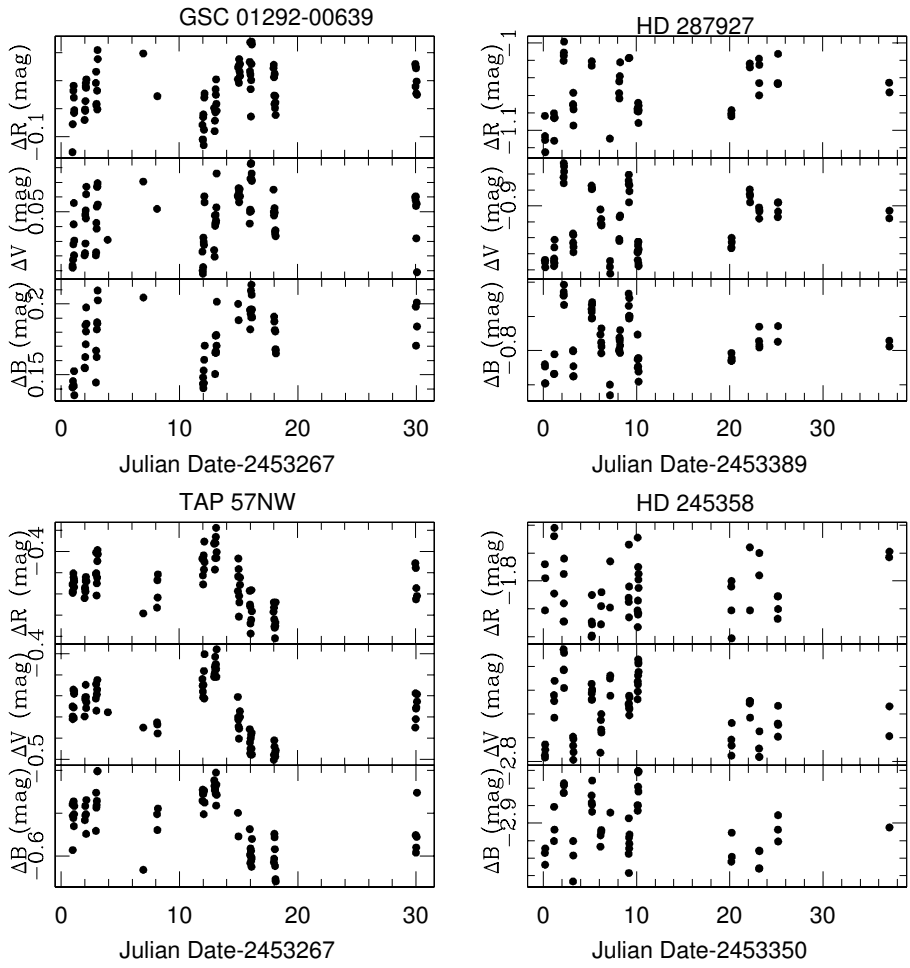

**Fig. .1** continued

**Table .2** Photometric amplitudes in the BVR and rotational periods of 12 WTTSs

| Stars | ΔB (mag) | ΔV (mag) | ΔR (mag) | periods (days) | (B-V) |
|---|---|---|---|---|---|
| [LH98 ]37 | 0.155 | 0.140 | 0.125 | 1.130 | 0.85 |
| [LH98 ]53 | 0.175 | 0.161 | 0.146 | 0.728 | 0.69 |
| HD 285372 | 0.268 | 0.205 | 0.153 | 0.287 | 1.07 |
| HD 283323 | 0.105 | 0.086 | 0.082 | 1.93 | 0.82 |
| HD 284503 | 0.191 | 0.150 | 0.148 | 0.741 | 0.65 |
| HD 285840 | 0.160 | 0.150 | 0.128 | 1.558 | 0.84 |
| HD 283716 | 0.079 | 0.077 | 0.065 | 1.48 | 0.94 |
| HD 282346 | 0.122 | 0.110 | 0.109 | 0.730 | 0.77 |
| GSC 01292-00639 | 0.080 | 0.080 | 0.080 | 0.919 | 0.93 |
| TAP 57NW | 0.162 | 0.132 | 0.096 | 9.12 | 1.12 |
| HD 287927 | 0.162 | 0.138 | 0.126 | 0.772 | 0.80 |
| HD 245358 | 0.126 | 0.093 | 0.085 | 0.736 | 0.76 |

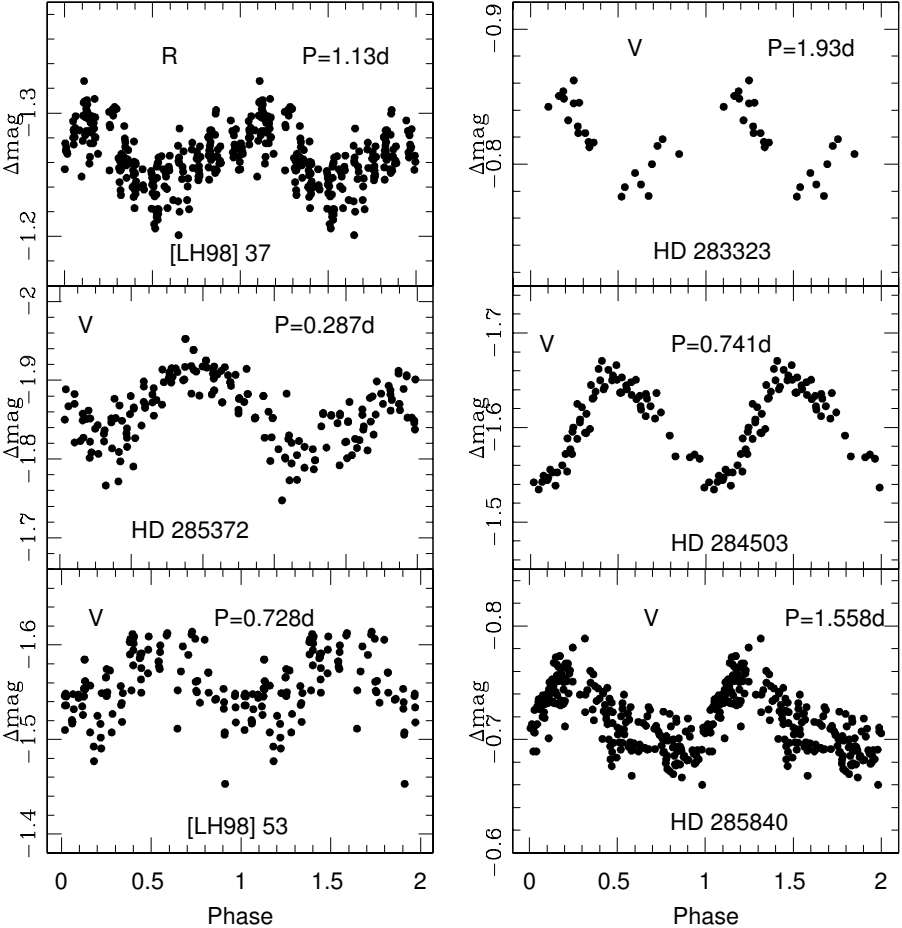

**Fig. .2** The phase-folded light curves of the 12 WTTSs

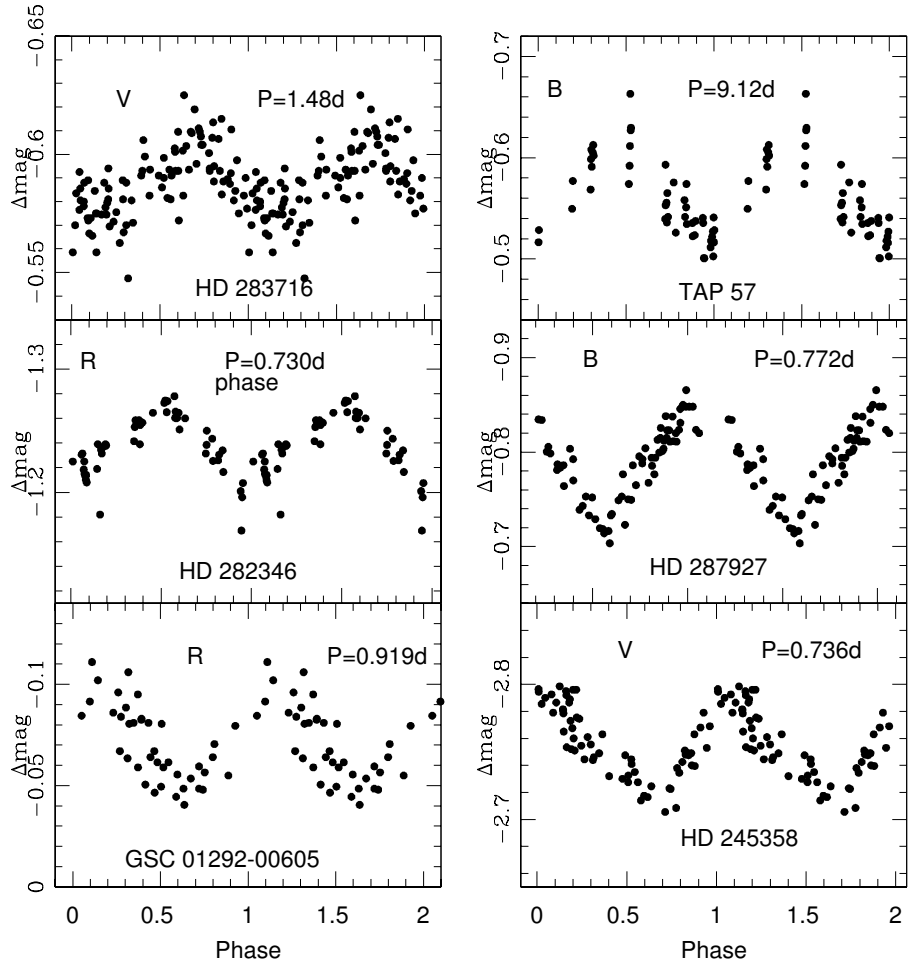

**Fig. 2** Continued.

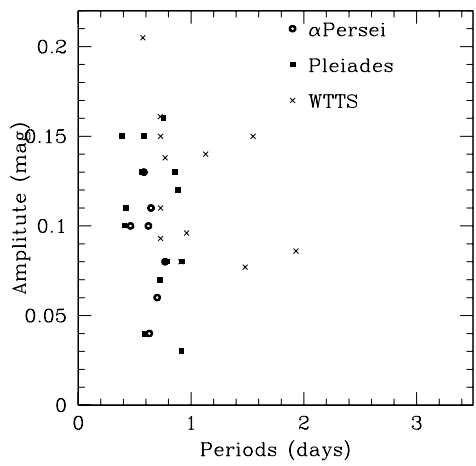

**Fig. 3** The rotational periods versus V light-curve amplitude

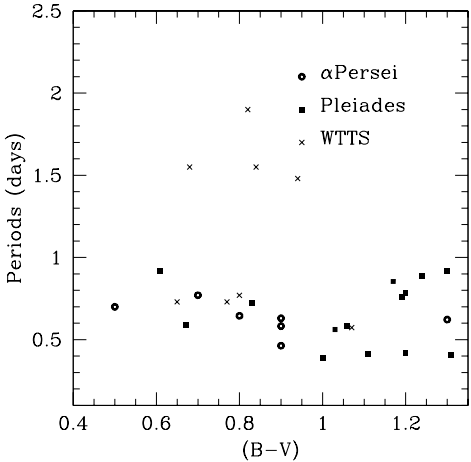

**Fig. .4** The rotational periods versus (B-V) color indices for stars belonging to the young open cluster $\alpha$ Persei (empty circles), Pleiades (filled square) and our sample of WTTS (crosses).

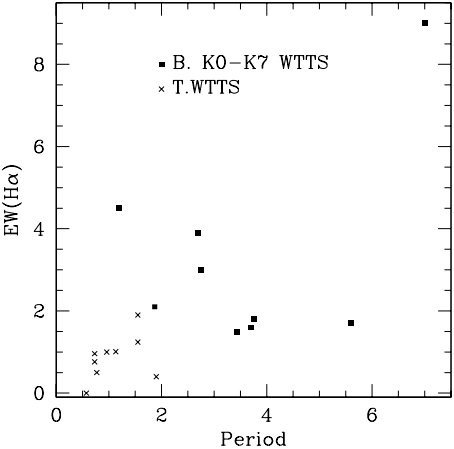

**Fig. .5** The rotational periods versus EW(H$_\alpha$) for stars belonging to the WTTSs (filled square) and our sample of WTTS (crosses).

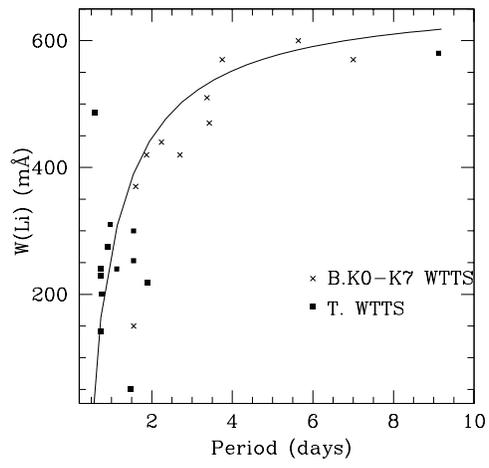

**Fig. .6** The rotational periods versus EW(Li) for stars belonging to the WTTSs (filled square) of Bouvier et al. ( 1993) and our sample of WTTS (crosses).